# AI Fairness: from Principles to Practice

**Arash Bateni, PhD**
Practice Lead, Accenture Responsible AI for HR
arash.bateni@accenture.com

**Matthew C. Chan**
PhD Candidate, Lancaster University, UK
m.c.chan@lancaster.ac.uk

**Ray Eitel-Porter**
Global Lead, Accenture Responsible AI
ray.eitel-porter@accenture.com

**ABSTRACT**

This paper summarizes and evaluates various approaches, methods, and techniques for pursuing fairness in artificial intelligence (AI) systems. It examines the merits and shortcomings of these measures and proposes practical guidelines for defining, measuring, and preventing bias in AI. In particular, it cautions against some of the simplistic, yet common, methods for evaluating bias in AI systems, and offers more sophisticated and effective alternatives. The paper also addresses widespread controversies and confusions in the field by providing a common language among different stakeholders of high-impact AI systems. It describes various trade-offs involving AI fairness, and provides practical recommendations for balancing them. It offers techniques for evaluating the costs and benefits of fairness targets, and defines the role of human judgment in setting these targets. This paper provides discussions and guidelines for AI practitioners, organization leaders, and policymakers, as well as various links to additional materials for a more technical audience. Numerous real-world examples are provided to clarify the concepts, challenges, and recommendations from a practical perspective.

## 1. INTRODUCTION

### 1.1 Structure of this Paper
This paper is structured in five sections. In the Introduction, we use a hypothetical case study to explore important principles of AI fairness and to define the paper's scope. The following two sections provide in-depth discussion of two broad notions of AI fairness: the treatment of individuals vs. the model's outcomes. The fourth section describes some of the inherent trade-offs and practical challenges involved in achieving AI fairness. In the final section, we provide practical guidelines for overcoming such challenges. At the end, we also include an additional list of references and resources for a technical audience.

### 1.2 Scope
Increasingly, AI is being used in a variety of sensitive applications that affect human lives. These range from recruitment and compensation, to consumer lending, to healthcare and criminal justice. This proliferation of AI use cases has led to the novel and emerging field of Responsible AI (RAI) [i] – the practice of designing, building, and deploying AI in a manner that empowers organizations, while treating people fairly [2]. RAI principles enable companies to foster trust in their AI models and scale these models with confidence [ii, iii]. The field of RAI encompasses various aspects of AI systems including accuracy, interpretability, explainability, accountability, privacy, and fairness. This paper focuses on AI fairness, one of the pillars of RAI.

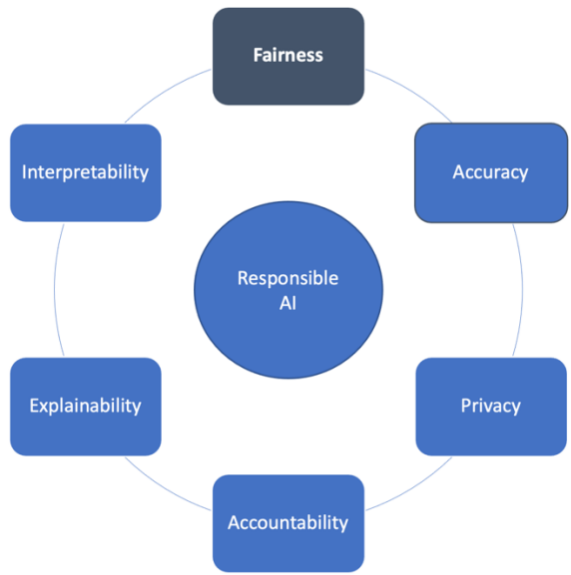

**Figure 1**: Various aspects of Responsible AI

Traditionally, fairness is regarded as an abstract concept and its application is often dependent upon context. However, to apply fairness consistently throughout the full universe of AI applications, the concept not only must be defined clearly, but also quantified, measured, and verified. In RAI, the quantifiable level of unfairness is generally referred to as bias [1]. The topic of AI fairness covers the prevention, measurement, and removal of bias. This paper is primarily focused on prevention and measurement, with some reference to removal techniques.

This paper describes common notions of fairness, as well as various methods used to define and quantify bias in AI systems. It examines the merits and drawbacks of different fairness definitions, and explains the challenges in establishing fairness in probabilistic processes. The paper emphasizes the role of human judgment in ensuring fairness, and clarifies how AI output can itself help inform fairness decisions. It also cautions against common, yet simplistic, practices for evaluating AI fairness. Instead, it proposes practical guidelines for more sophisticated and effective approaches toward fairness, which can be employed by AI practitioners, organization leaders, and policymakers.

**1.3 Case Study: The Impact of Familiarity Bias on Fairness**

Consider the hypothetical example of a university admissions office aiming to identify applicants with the highest chance of succeeding in its graduate program. The office carefully evaluates both domestic and international applicants based on their undergraduate performance. However, the university does not have the same level of familiarity with all applications. Because it has historically received a higher rate of domestic students, the admissions team is more familiar with these candidates' applications, profiles, and performance.

Since the university is less familiar with international applicants, it is more likely to make poor decisions about admitting them. These poor decisions appear in two forms: accepting unqualified applicants (false positives) or rejecting qualified ones (false negatives). Both of these errors occur at higher rates for international applicants compared to their domestic counterparts.

The admissions office has established a rigorous program to track student performance, create a feedback loop, and learn from its mistakes. The performance tracking program shows that a higher percentage of international students fail to complete their graduate study at the university. This is due to the fact that the office has disproportionately admitted unqualified international applicants and rejected qualified ones. However, based on its feedback loop, the admissions office will (incorrectly) conclude that the pool of international applicants is of lower quality (even when its true quality is at the same level as domestic applicants).

---

[1] Technically, bias may exist in various forms of data, algorithms, and procedures. The bias is referred to as unfairness when it impacts individuals. Furthermore, unfairness is not limited to bias. Poor predictive accuracy is unfair to qualified individuals, even when it does not cause bias (e.g., random hiring or lending).

In other words, the admissions team's unfamiliarity with international applications not only leads to poor admissions decisions (a form of error), but also creates a misleading feedback loop indicating that less familiar applicants are lower quality (a form of bias) [2]. This phenomenon can be called "familiarity bias" [3].

**(a)**

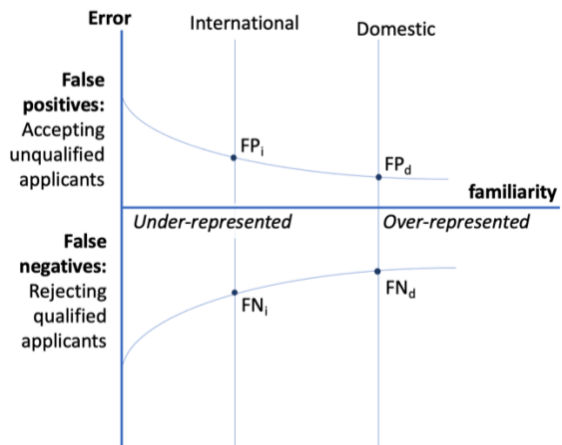

**(b)**

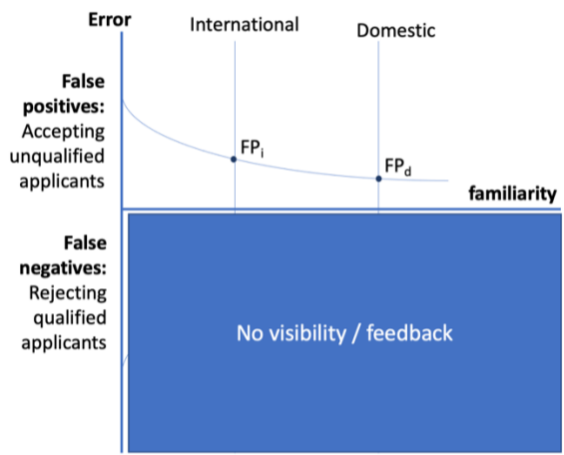

**(c)**

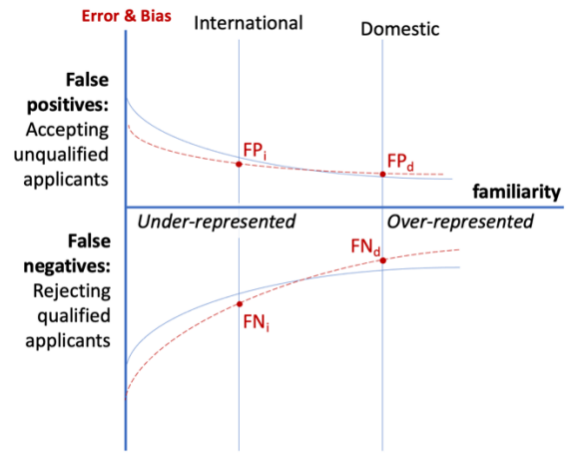

**Figure 2:** Illustrations of familiarity bias, assuming that domestic and international applicants have identical qualifications.

(a) Familiarity with domestic applications leads to lower error rates compared to international applications. However, this initial error can result in similar proportions of false positives and false negatives ($FP_i / FN_i \cong FP_d / FN_d$). This suggests that while the admissions office makes more mistakes for one group, it is not favoring another one.
(b) Feedback data is only collected on admitted applicants. This means that the admissions office has no visibility into the potential performance of the rejected individuals. The feedback shows that domestic students achieve a higher success rate (i.e., fewer false positives). This leads to the admissions office incorrectly concluding that domestic students are of a higher quality.
(c) Based on the feedback, admissions decisions are adjusted to reject international applicants at a higher rate, aiming to narrow the gap in false positives ($FP_i$ and $FP_d$). The adjustment has a corresponding effect on the false negative curve (new dashed line), manifested in the form of a higher rejection rate of qualified international applicants.

---

[2] As described later in this paper, a high level of error in decision-making is a source of unfairness in itself, as it leads to the rejection of qualified individuals and the acceptance of unqualified ones. This is the case even when the error does not cause bias among groups (e.g., in a randomized selection process).

[3] Note the distinction between familiarity bias (illustrated in Figure 2-c) and commonly known representation and measurement biases (illustrated in Figure 2-a).

Familiarity bias occurs when two conditions are met: a) there is imbalance in input data [4] (e.g., a smaller number of international applications), and b) there is incomplete feedback data [5] (e.g., the admissions team does not collect feedback data on rejected candidates, so it does not discover the higher rejection rate of qualified international applicants). The latter condition indicates that feedback data is not representative of the entire population, and the exclusion of individuals (i.e., rejected candidates) from this data set depends on the selection process itself. This flaw means that the effects of the initial process errors are reproduced, thereby perpetuating the familiarity bias.

The underlying conditions behind familiarity bias are not rare at all. Varying levels of familiarity toward given subsets or groups of the population are common in many environments. This phenomenon may be caused by historical discrimination or any other reason leading to the under-representation of a group. The bias can also be caused by the smaller size of minority groups in the overall population.

The second condition for familiarity bias is even more common. In fact, in most real-world processes it is inevitable that feedback data is not collected from rejected individuals. In critical processes, such as recruitment, admissions, and consumer lending, the decision-makers only have visibility into the performance of the individuals they accept (see Figure 2-b).

Familiarity bias can therefore lead to a disproportionate rejection of equally qualified individuals from under-represented groups. Significantly, this process does not assume any form of conscious or unconscious human bias. When the underlying conditions described above are met, familiarity bias arises even when people are perfectly objective and unbiased.

**1.4 Fairness in AI vs. Manual Systems**

As the above hypothetical example demonstrates, familiarity bias can arise and perpetuate in manual processes as well as data-driven AI systems. This observation can also be applied more widely. The types of bias, and the methods through which they enter into given processes, are similar across AI and manual systems. Therefore, many of the topics discussed here in the context of AI fairness are readily applicable to non-AI systems. This includes notions of fairness, the mutual incompatibility of fairness definitions, and the trade-offs between fairness and other social concerns.

AI can in fact create an opportunity to uncover previously unknown biases, and provide methods and techniques to measure, monitor, and control them in ways not possible before. It can empower organizations to make informed decisions about fairness targets, by revealing the inherent trade-offs involved in balancing fairness with accuracy, privacy and other factors.

Having said that, AI has also created unique challenges, because when AI models are being trained, the feedback process is automatic. Any issues in the training data, such as those described in the familiarity bias example, quickly translate into outcomes. Because AI learns so efficiently, it risks scaling and magnifying biases to a greater degree than manual systems.

The use of AI may suggest that data-driven outcomes are inherently more objective and reliable than human decision-making. This is only partly true. AI systems formalize the decision-making process, and they tend to be more objective and less prone to human prejudice [iv]. However, they are not immune from bias. For instance:
- Training data is often imperfect and may be tainted by historical biases.

---

[4] The imbalance in data can be rooted in the number or quality of observations.

[5] A similar phenomenon is referred to as "survivorship bias". It means concentrating on the people or things that made it past some selection process and overlooking those that did not, typically because of the lack of visibility.
https://towardsdatascience.com/survivorship-bias-in-data-science-and-machine-learning-4581419b3bca. Accessed 12 February 2022.

- Design decisions, such as the choice of attributes and features fed into a model, can significantly impact the balance of outcomes.
- Given the widespread availability of model libraries and access to large amount of public data, novice practitioners can build and launch AI solutions quickly and without proper testing. They may not fully realize the subtleties of the domain and the implications of deploying AI-based solutions.

When it comes to fairness, one unique aspect of AI concerns agency and accountability. Generally, the designer of an AI model has limited control over its behavior. They do not set the rules as to how the tool should perform in each situation. Instead, AI learns such rules from patterns in its training data, which may be created well after the model is designed and launched. In many situations, the designer cannot predict the outcomes that their AI system will generate. Depending on the methods used, they may not be able to explain precisely why the model reaches a certain outcome. This characteristic of AI is different from any other tool, where the designer has complete control over the product's features and characteristics and can therefore take full responsibility for its performance.

Similarly, in most cases, the end user of an AI system has limited understanding of or control over how the tool arrived at a result, so they cannot take full responsibility for it. Again, this is in contrast with other tools, where the user can be held responsible if the tool makes unfair decisions.

AI systems have significant amounts of agency to learn, evolve, and make predictions/decisions independently. When the outcome is unexpected or unfair, it can be difficult to determine who is accountable. All players ranging from AI designers to end users may act responsibly and with good intentions, yet achieve unfair outcomes. The topic of AI accountability is not the focus of this paper; however, it is a critically important pillar of RAI [v].

**1.5 Positive vs. Neutral Approaches Toward Fairness**

Any observed pattern in the outcomes of an AI system may be rooted in the system itself or in upstream factors, such as a) inherent differences between groups (e.g., women being safer drivers than men [vi]), b) the preferences of groups (e.g., younger employees having a greater tolerance for long commutes [vii]), or c) historical discrimination or biases (e.g., historical segregation and racism against African Americans).

An AI system in turn consists of inputs and models, and is influenced by various considerations such as the training data, the selection of driving factors, the design of algorithms, the optimization objectives, and the tuning and validation process.

AI fairness can rely on two distinctly different approaches. The "neutral" approach assumes that individuals with the same merit or qualifications should receive the same outcome (e.g., scores), regardless of which group they fall into. To achieve this neutrality, AI systems should rely solely on each individual's characteristics related to the model's purpose, disregarding other upstream factors.

The "positive" approach to fairness, on the other hand, argues that the observed qualifications of groups within the population may be partly influenced by "unjust" factors, such as historical discrimination. Therefore, the AI system should treat groups differently to account for such factors. For instance, this could mean that historically disadvantaged groups receive higher scores for the same level of qualifications.

In practice, embedding positive intervention in AI systems is challenging. As mentioned above, the differences between groups may be the result of any number of upstream factors (inherent behaviors, preferences, or historical discrimination). To design AI systems that reverse the effect of unjust components of the upstream factors, one would need not only to identify these components, but also to isolate and quantify their impact on differences between the groups. There is no consensus on a methodological approach to achieve this.

Furthermore, differences between the population groups are specific to a time and location. This means that organizations would need to design, deploy, and maintain various versions of an AI system for different environments. This would in turn require a complex governance structure to determine which groups should be favored by AI, to what degree, in which area, and for what timeframe. Modifying AI systems to favor disadvantaged groups also poses major challenges in the interpretation, usability and explainability of the results, as the outcomes for different groups, timeframes or regions may not be comparable.

In addition to these operational challenges, the positive approach to fairness involves ethical trade-offs [vi, vii, viii]. Consider an idealized scenario where the impact of historical discrimination on certain groups can be perfectly identified, quantified, and reversed by AI. Such adjustments would align the groups at aggregate level (e.g., by ensuring a similar mean or median across groups). However, to achieve this, the adjustments also negatively impact disadvantaged individuals at the lower end of a generally privileged group. For this reason, positive interventions at group level may be perceived as being unfair at individual level.

Finally, adopting the positive approach to fairness requires that group attributes directly enter the models and influence outcomes. However, this is problematic in practice, because favoring one group over another and making decisions based on protected attributes, such as age, gender, or race, is outlawed in many jurisdictions [ix, x, xi, xii].

Due to these operational, ethical, and legal limitations, positive interventions are not suitable for inclusion in AI systems. Instead, AI should be designed to be neutral – producing outcomes that accurately reflect the observed merit and qualifications of individuals, without any bias. Positive interventions, when required and permitted, may then be applied on top of given AI outcomes. End users can decide how the scores should be interpreted, used, or adjusted in accordance with their organization's objectives and policies. Capturing the history of these adjustments enables AI designers to isolate and analyze the magnitude and impact of the positive interventions.

This paper is written from the perspective of the neutral approach toward fairness, which assumes that no group, sub-group, or individual should be favored by AI. This approach leads to two broad notions of fairness: equal treatment and equal outcome (extensively discussed in legal literature as "disparate treatment" and "disparate impact" [xiii, xiv, xv]). The following sections of the paper discuss each of these approaches in detail, along with their pros and cons, illustrative examples, and practical guidelines.

## 2. THE "EQUAL TREATMENT" NOTION OF FAIRNESS

The "equal treatment" concept of fairness in AI focuses on the process of generating outcomes, rather than the outcomes themselves. It aims to ensure that AI systems treat everyone in the same manner, without favoring some individuals over others. More specifically, equal treatment approaches hold that AI should be blind or indifferent to any attribute or feature that is not related to the model's purpose. This includes demographic attributes (e.g., age, gender, ethnicity, etc.), for example, because in most practical cases, they should not influence the model's outcome.

In principle, the journey toward equal treatment should cover all aspects of AI, ranging from sources of input data to feature selection, to the choice of algorithms and model hyperparameters. In the rest of this section, we focus on several measures which would largely influence the treatment of groups and individuals: a) evaluating the data for bias, b) evaluating the data for balanced representation, and c) controlling the features in the data [6].

---

[6] Considering AI input as a matrix, these measures address rows (evaluation for representations), columns (controlling features), and cells (evaluation for bias).

## 2.1 Evaluating the Data for Bias

If an AI is trained on intrinsically biased data, the bias will be translated into its outcomes. This process is known as "bias in, bias out". Biases within training data can be due to unfair processes or practices, conscious or unconscious bias by human decision-makers, or in cases like familiarity bias, because of the unbalanced representation of groups.

The risk of bias within training data is significantly higher when the data is influenced by the AI outcome itself. Consider attrition prediction models as an example. These models are trained on data from actual (employee or customer) attrition, which can be assumed to be accurate, unbiased, and independent of the model's predictions. However, when model predictions are used to prevent attrition, then subsequent data becomes influenced by past predictions and preventive actions. This creates a self-reinforcing feedback loop that muddies the training data.

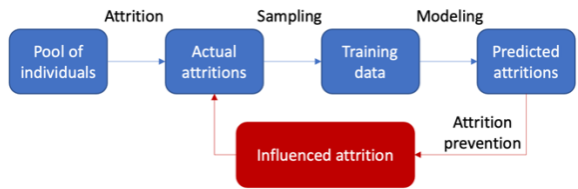

**Figure 3:** Illustration of a self-reinforcing feedback loop in attrition prediction systems

For another example, consider an application of AI in auditing, where an AI system predicts the likelihood of mistakes by individuals, and hence provides targeted recommendations for future audits. The AI should be trained on actual mistakes by individuals, and data obtained through an independent process (preferably randomized audits). However, if the training data consists of mistakes uncovered by past AI-driven audits, then it is prone to the self-reinforcing feedback loop and an elevated risk of bias. The risks posed by this type of feedback loop in the area of predictive policing are well-documented. They clearly illustrate the negative impact that such loops can have on groups and individuals [xvi, xvii].

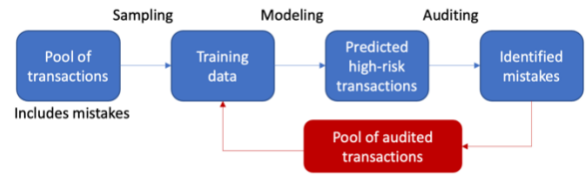

**Figure 4**: Illustration of self-reinforcing feedback loop in audit prediction systems

These feedback loop examples highlight the difficulty in identifying the risk of bias within training data. In most cases, identification cannot be achieved through technical means alone, because it requires an understanding of the data sources and the underlying process used to create the data.

It may be tempting to rely on the disproportional representation of groups within training data as an indication of bias. However, to do so would be ineffective and potentially misleading. As described below in section 3 on "equal outcome", having varying degrees of group representation within training data may be justified based on merit and qualifications, and therefore by itself should not be interpreted as evidence of bias. Despite this, having significant under-representation of groups within training data, even if it is justified, may lead to other issues as described in the next section on evaluating the representation of groups.

Most of the techniques available to check for bias within training data are context-based. This means that they compare two data sets that have been created under different conditions. Examples include comparing data from different sources, comparing subsets of data from different timeframes, and comparing data obtained from manual vs. AI systems. These checks can also include various types of back testing, i.e., replicating portions of the data through manual or AI processes. This might mean, for instance, shuffling data to examine whether expected outcomes are independent of sensitive attributes [xviii, xix]. If the results obtained under different conditions show gaps between different groups, it would indicate that these gaps stem from the conditions under which the data was created.

A more advanced technique that may be applicable to some use cases is replicating portions of training data using supervised and unsupervised [xx] models in parallel. Consider an example of a model that matches people's resumés to job descriptions. An unsupervised approach could involve a model that analyzes the text within documents to identify overlaps between applicants' resumés and the job description in question, without using any training data. A supervised approach could involve training the model to learn from known hiring outcomes. The training process would determine which sections of the resumé are better predictors of the actual hiring outcome. A supervised model would likely yield more accurate matches than an unsupervised model, but it would also be prone to the risk of bias in the training data (e.g., human bias in historical hiring outcomes). In this case, comparing the results of supervised and unsupervised models could help isolate and uncover the potential impact of bias within the training data.

## 2.2 Evaluating the Data for Balanced Representation

As outlined above, familiarity bias arises when there are varying degrees of familiarity toward subsets of the population. In AI systems, this occurs when groups do not have similar (or sufficient) representation in the training data. Generally, fewer data points from under-represented groups will correspond to reduced familiarity and higher error rates, as described in the case of university admissions. Such errors can be further magnified in AI systems, because algorithms will optimize for groups with greater representation at the expense of minority groups [xxi]. The AI system expects the behavior of the dominant group from the under-represented groups, ignoring their differences, which further increases the error rates for these groups. These errors translate into biased feedback loops and the disproportionate rejection of equally qualified individuals from the under-represented groups.

In practice, there are other operational limitations that further work against these minority groups. For example, widely used adverse impact tests (see section 3.1 below for more detail) are often limited to large and well-defined groups. This is because data collection for these groups is less expensive, sufficient data points are more readily available for statistical analysis, and there is a greater need for testing since there are more significant operational, legal, or reputational risks attached to negative outcomes for these groups. Additionally, impact tests require accurately labelled demographic data, which is more readily available for larger groups (e.g., gender) and more challenging to acquire for smaller groups (e.g., most ethnicities). Because of these limitations, the under-represented groups that are mostly impacted by the familiarity bias are also omitted from adverse impact testing, which aims to detect such biases. As a result, the magnitude of the bias remains unknown, and no actions are taken to mitigate it.

For these reasons, outcome tests, such as those assessing adverse impact, are ineffective or inapplicable when addressing familiarity bias in practice. Instead, AI designers should aim for equal treatment by ensuring that all subsets of the population are sufficiently represented in the training data. There are a variety of methods and techniques available to analyze imbalance within training data [e.g., xxii, xxiii, xxiv, xxv, xxvi, etc.].

However, the evaluation of training data is not a purely technical activity. Training data should be a sample that is representative of the entire population. As in any sampling, the process should be examined to ensure it leads to correct representation of all population segments. This process should be randomized, particularly for sensitive use cases. For instance, face detection models employed by law enforcement should be trained and validated on a set of images that have been randomly selected from the entire population.

Self-reinforcing feedback loops also need to be addressed – particularly where sampling and group representation are concerned. In such cases, training data must be scrutinized before it is used in AI modeling. Otherwise, there is a risk that familiarity bias from existing human or AI processes is magnified by the further exclusion of under-represented groups from the training data, exacerbating unfair outcomes.

If it is not possible to establish balanced training data, then AI designers can use various bias removal techniques to amplify the effect of under-represented groups [xxvii]. Bias removal techniques generally tweak the data, the models, or the end results to achieve desired patterns [7]. While these techniques may reduce the effect of bias, they often have negative side effects, such as diminishing AI's predictive accuracy [xxviii]. Therefore, for sensitive applications (e.g., employing AI in criminal sentencing), we recommend obtaining additional observations to make the training data representative of the population, rather than using bias mitigation techniques.

Achieving unbiased and balanced representation in training data should be considered a critical element of AI development for several reasons:
- It is an important step toward the principle of equal treatment of individuals.
- It not only reduces the risk of bias, but also improves the solution's predictive accuracy (whereas many other methods for achieving fairness come at the expense of accuracy).
- It addresses the root cause of bias and fixes the flaws in the data at the source, before the data is used by the AI system.
- It does not require the designer to tweak the model or algorithms to remove bias.
- The evaluation of training data and its sources is model-agnostic; the data only needs to be examined once and then it can be used for various applications.

**2.3 Controlling the Features in the Data**
When AI predicts an outcome for an individual, it is essentially relying on the historical outcomes of people with similar attributes. For example, an AI will reject a new customer's loan application if the data shows that applicants with similar attributes did not repay their loans. Therefore, both the accuracy and fairness of the predictions directly depend on how the similarity between individuals is measured. This measurement is achieved during a critical step in model development called feature selection, which determines which attributes, factors, or variables are passed to the AI system during its training process.

Feature selection can directly impact AI fairness and should therefore be considered a vital part of model development, particularly for sensitive use cases. For the sake of fairness, the similarity of individuals (and consequently the predicted outcome) should not be determined by attributes such as age, race, or gender. Instead, features that describe the merits and qualifications of individuals in relation to the desired outcome should be used.

Generally for AI development, all data is perceived as potentially valuable, and therefore may be considered for use. For low-impact use cases, models can include many different features to help maximize their predictive accuracy. For instance, in digital advertising, a broad range of attributes including age, gender, and sexual orientation may be considered to predict behaviors [8, 9, 10].

---

[7] Referred to as pre-processing, in-processing, or post-processing techniques.

[8] Equal opportunity laws for housing, employment or credit may be extended to marketing or advertising of such products or services.

[9] There has been growing regulatory effort to limit the use of sensitive data even for low-impact applications. For instance, there has been new limitation on the use of intimate browsing history for digital advertising. These efforts are generally motivated by privacy rather than fairness. [see Cookies policy. European Commission. https://ec.europa.eu/info/cookies_en#thirdpartycookies. Accessed 14 February 2022, and Ultimate Guide to EU Cookie Laws. Privacy Policies. https://www.privacypolicies.com/blog/eu-cookie-law/ (2021). Accessed 14 February 2022.]

[10] In some cases, targeting advertising to certain demographics has caused major ethical concerns [see Tobacco is a social justice issue: Racial and ethnic minorities. Truth Initiative. https://truthinitiative.org/research-

On the other hand, for high-impact application areas, feature selection depends on fairness as well predictive accuracy. Controlling or limiting input data for fairness reasons may compromise accuracy (see section 4.2 below for more information on trade-offs between fairness and accuracy) [xxix]. However, this is a recommended practice for sensitive areas, such as recruitment and mortgage lending, where fairness is a priority.

Feature selection from a fairness perspective cannot be an entirely automated process. It requires human judgment to examine the role of each feature, determine the desired trade-off between accuracy and fairness, and take into account various operational, legal, and reputational considerations.

It is generally accepted that the demographic attributes, as well as their proxies, should be excluded from AI systems. However, identifying the proxies is not straightforward. It is common practice to treat any feature with a strong correlation with demographic attributes as a proxy. In our view, this is incorrect. To clarify this point, consider the example of HR, where certain skills may correlate with age or gender, or in consumer lending, where income may correlate with race. In each of these cases, the correlated features are eligible for use by AI tools, as they describe individuals' observed merit and qualifications in relation to the target outcome.

Therefore, the AI designer should consider the features' relevance to the AI's purpose. If the features are directly relevant to the system's purpose, they should not be removed, even though they correlate with demographic attributes. In recruitment, examples of such features include skills, education, and experience. In credit risk modeling, they include financial metrics such as income, debt, and credit utilization.

To further clarify this point, consider "postal code" as a feature. In countries with a history of racial segregation, such as the U.S. and South Africa, postal codes often correlate with race. Therefore, this feature should be excluded from many AI applications such as credit risk modeling, where it does not directly relate to the tool's purpose (i.e., there is no reason to believe that a postal code by itself may influence an individual's credit risk). However, postal code may be used in other AI applications, such as flood or fire risk models. In these cases, the feature is directly relevant to the AI's purpose and therefore it is eligible to be used, its correlation to race notwithstanding.

While the above guidelines should be adopted as common practice in AI development, different approaches may be taken for more advanced AI systems, where deliberately including certain demographic attributes would improve predictive accuracy and reduce bias.

Consider the case of credit scoring models. Credit history length is commonly used as a feature in such models, where fewer years of history translates into a lower credit score. However, this feature carries different meanings for different subsets of a population. While a short credit history would carry negative implications for an adult who has spent their entire life in that country, it may only represent the number of years that an immigrant has lived there. In this case, adding the years of residency as a new feature to the model would benefit predictive accuracy while mitigating bias.

This approach may also be relevant to other use cases. For example, in employee attrition prediction, commute times may be considered for inclusion in the model, because this feature may impact employees' perceptions of their working lives. However, this feature may not impact all individuals in the same manner. For instance, younger employees may have a higher tolerance level for time spent away from home. Therefore, including attributes such as age in an attrition

---

resources/targeted-communities/tobacco-social-justice-issue-racial-and-ethnic-minorities (2017). Accessed 14 February 2022, and Apollonio, D.E.,

Malone, R.E.: Marketing to the marginalized: tobacco industry targeting of the homeless and mentally ill. Tobacco Control 2005;14:409-415]

prediction model may both improve accuracy and reduce bias in outcomes.

There are other advanced modeling techniques that require the inclusion of demographic attributes. These models rely on demographics to achieve certain definitions of fairness in outcomes (e.g., similar error rates among groups). This is achieved by modifying the core algorithm to co-optimize accuracy and fairness at the same time. In this case, the definition of fairness is built into the algorithm itself. This offers a robust approach which avoids suboptimal combinations of accuracy and fairness (see section 4.2 below on the Pareto frontier for more details).

Generally, any method that allows demographic attributes as an input should be used cautiously. AI designers should be equipped to fully examine and analyze the impact of each attribute on the model's accuracy and bias. The benchmark data that is used to measure accuracy and fairness should itself be reliable, accurate, and unbiased. The sources and the process of data collection should be scrutinized. Furthermore, training data should be checked to ensure a proper representation of the entire population. These safeguards should be adopted with greater attentiveness when demographic attributes are allowed into the model, since the model will no longer be blind to these attributes and this may result in the principle of equal treatment being violated.

Finally, it should be noted that in some jurisdictions, the use of demographic attributes is not permitted, regardless of the impact on accuracy or bias. For example, the European Union prohibits the use of gender in calculating car insurance premiums [xxx], despite evidence suggesting that women tend to be safer drivers [xxxi]. In some other cases, it may be considered morally unacceptable to use demographic attributes, even if they are functionally and statistically relevant [xxxii]. For example, racial profiling by law enforcement [xxxiii, xxxiv] is unacceptable on moral grounds, regardless of whether it has a functional outcome.

## 2.4 Considerations Around Equal Treatment

Some people believe that equal treatment approaches, and in particular controlling features in the data, are ineffective. Their argument is that when AI knows everything about individuals, it can infer their demographic attributes even when these attributes are not directly provided to the model [xxxv, xxxvi]. In principle, this is true [11]. However, this criticism is irrelevant when models use only limited features that are directly relevant to the solution's purpose – an approach recommended for high-impact applications. This method is different to general-purpose AI, where a model considers all available data to be potentially useful.

Furthermore, correlations between outcomes and demographic groups should not be confused with a scenario where demographic attributes are inferred by the model. Technically, such correlations in the outcomes may be simply a reflection of the corresponding correlations between the relevant features and the demographic groups. As described earlier, the use of features correlated with demographics is common, and may be justified when these features reflect attributes or qualifications relevant to the tool's purpose [12]. For example, postal code is a relevant and justified feature for fire and flood risk models (but not for credit risk models). In some places, this feature correlates with ethnicity or age. This would be reflected in an outcome which will also correlate with these attributes. It should be noted that equal treatment, or controlling features in the data, does not aim to enforce equality (e.g., statistical parity) across groups [13] (for more information, see section 3.2 below on conditional statistical parity).

---

[11] For instance, there are commercial AI solutions for inferring demographic attributes such as gender, ethnicity, and country of origin, by using first name, last name, and postal code [see Namsor, name checker for gender, origin and ethnicity classification. https://namsor.app/]

[12] Certain use cases will involve correlated features that are not representative of individual's merit. In facial recognition, for example, features such as skin tone and hair length correlate with gender and ethnicity [see Wehrli, S., Hertweck, C., Amirian, M. et al. Bias, awareness, and ignorance in deep-learning-based face recognition. AI Ethics (2021). https://doi.org/10.1007/s43681-021-00108-6]. For such use cases, feature control would not be effective and the focus should be on other measures, such as evaluating the data for balanced representation.

The primary limitation of the equal treatment notion is that it is inherently open-ended and does not have a verifiable target. It consists of various quantitative and qualitative measures within AI design and development. However, the goal of treating everyone the same cannot be tested and validated quantitatively. In other words, although the measures described here provide important benefits, they cannot guarantee that equal treatment is in fact achieved. To mitigate this limitation, we recommend complementing the measures of equal treatment with quantitative tests on model outcomes (e.g., conditional statistical parity or equal accuracy). This is described in section 5 below on practical guidelines.

Despite this limitation, equal treatment is a powerful approach to pursuing fairness. Many of the measures described here can be implemented regardless of group definitions, and can therefore benefit various groups, sub-groups, and individuals. Equal treatment mitigates bias by addressing its root cause at the source. It aims for fairness by design, rather than seeking it after the fact (e.g., in statistical parity). The focus is on the input, training data, and feature selection – independent of the modeling techniques, outcome structure, or tool's purpose. As a result, these methods are applicable to a wide range of AI solutions. For these reasons, the benefits of this notion far outweigh its limitation. We recommend regarding equal treatment as a foundational principle in AI fairness.

## 3. THE "EQUAL OUTCOME" NOTION OF FAIRNESS

The notion of "equal outcomes" aims to establish specific patterns in AI outcomes that are perceived to be fair. Inherently, this approach is more quantitative and verifiable than the notion of equal treatment. Many of the tests for equal outcomes can be performed regardless of the processes or models used for generating the results.

In principle, the notion of equal outcomes can be applied to individuals or groups, which means that the expected or fair patterns in AI outcomes should be defined at the individual or group level. In practice, most existing techniques for ensuring equal outcomes are developed and deployed at the group level only (i.e., to assess fairness across pre-defined groups, rather than between individuals). Group definitions can vary case by case. In some jurisdictions, there are legal protections banning discrimination against certain groups, such as people of a certain age, gender, ethnicity, or religion, or those who have a disability [see xxxvii, xxxviii, xxxix, xl, xli, xlii, xliii, xliv]. Consequently, many organizations test their models for bias against these groups.

### 3.1 Statistical Parity
Statistical parity tests (also referred to as demographic parity, adverse impact or disparate impact tests) requires that AI outcomes are equally distributed across groups. For this purpose, the results are aggregated and compared across groups, typically using statistical tests. The tests can be applied directly to the numerical scores generated by an AI, or on the categorical decisions derived from the scores (e.g., whether the score is higher than an acceptance threshold). Additionally, the decisions can be tested as counts or proportions e.g., how many or what percentage of the individuals from each group are accepted.

To conduct statistical parity tests, an organization must set pre-defined tolerances for differences between groups. For example, some U.S. federal agencies follow a guideline stating that the selection rate for any given group should not be less than 80% of the group with the highest selection rate. This is referred to as the "four-fifths rule" [xlv, 14]. These tests are typically conducted according to regulations.

---

[13] Some confusion in the field arises from the expectation that controlling features in the data should "guarantee statistical parity" [see Dwork, C. et al.: Fairness through awareness. ACM Digital Library (2012). https://dl.acm.org/doi/10.1145/2090236.2090255]. For more information, see section 4.1 below on the mutual incompatibility of fairness.

Statistical parity has certain advantages. It is the simplest test of fairness, as it requires only a limited amount of information (i.e., AI outputs, which are almost always available). Therefore, it generally applies to a vast range of AI solutions (as well as manual procedures). It does not require scrutinizing the input data, the model, or the underlying process or assumptions for obtaining the results. It is purely quantitative and objective, and does not require human judgment. Therefore, the results can be easily interpreted by audiences with varying levels of expertise.

Due to its simplicity, statistical parity is the most widely used test of fairness. It is also accepted and recommended by policymakers and regulators. However, it also suffers from several weaknesses:

- **Limited insight into the fairness of AI systems**. Statistical parity focuses on one particular definition of fairness (i.e., equal distribution in results), and evaluates it for certain, predefined groups at aggregate level. There is limited flexibility around the definition of these groups, as designers require reliable data on demographic attributes such as age and gender. In practice, this approach cannot be applied to any group where attributes are not known (e.g., religious beliefs) or not available for the analysis (e.g., race or ethnicity in most cases). In practice, the test is only applied to a small subset of the legally protected groups (usually gender), ignoring other groups (for more information, see section 2.2 above on the operational limitations of evaluating the representation of minorities in training data).

- **Possible negative impacts on disadvantaged groups**. Although statistical parity tests fail to cover most disadvantaged groups, the methods of satisfying them may negatively impact these groups, because the tests do not consider how the equal outcomes are achieved. If organizations optimize AI solutions to enforce the desired pattern for certain groups, other groups may achieve worse outcomes. This has the opposite effect of equal treatment measures (e.g., scrutinizing data sources, testing for self-reinforcing feedback loops, and controlling the features), which seek to benefit all individuals and subsets within the population.

- **Possible manipulation of AI models**. To ensure equal outcomes across all groups, an AI designer can tweak their model to favor one group over another, intentionally violating the principle of equal treatment. These manipulations involve accessing and using the group attributes, further violating equal treatment. Furthermore, any such manipulation will likely negatively impact the model's utility and accuracy (for further information, see section 4.2 below on the Pareto frontier). In practice, the narrow focus of statistical parity can undermine broader notions of fairness, negatively affecting individuals and groups that are not included in the test.

- **Failure to account for individuals' merit or qualification**. Statistical parity ignores the features and attributes of individuals altogether. The expectation of achieving equal outcomes is only plausible in one of the two scenarios: a) individuals' qualifications are irrelevant to the outcome, and b) the different groups have the same qualifications. The first scenario is rare and the second one is unrealistic. To demonstrate this, consider a hypothetical model that relies entirely on a single feature (e.g., using income to determine credit risk). For the model to generate the same outcomes (risk scores)

---

[14] There is also legal precedent in the U.S. where the four-fifths rule is rejected, and the expectation is for zero difference between groups.

https://caselaw.findlaw.com/us-6th-circuit/1073540.

across different groups, it is not sufficient for all groups to have identical average or median incomes. Rather, all groups must have the same distribution of income, because any variation (e.g., in skewness or the tails of the distribution) could result in a corresponding difference in the scores. Any realistic model would employ numerous features as inputs. Therefore, all these features must have the same distribution across all groups to justify the expectation of statistical parity. This is practically impossible. In fact, the test of statistical parity would be violated by a perfect model. Such a model would make no mistakes in predicting individuals' performance. Yet it would not generate equal results when the features (merits) are not the same.

For the reasons described above, the test of statistical parity (or adverse impact) is flawed and ineffective for the evaluation of fairness [xlvi]. Despite these shortcomings, the simplicity and apparent interpretability of the approach means that it has been widely adopted by regulators and policymakers in many jurisdictions. In fact, it is widely relied on by practitioners and organization leaders as the primary test of AI fairness.

Allowing a pre-set tolerance for differences in outcome (e.g., the four-fifths rule [xlvii]) may be an attempt to address the flaws of statistical parity. While allowing such tolerances would make the equal outcome approach more realistic, the limits are arbitrary and not rooted in a scientific approach or the actual qualifications of the groups in question (for more information, see section 4.2 below on the Pareto frontier). The allowed tolerance may be too high or low for a given situation, making the result of the fairness test similarly arbitrary.

To overcome the inherent flaws in the statistical parity approach, fairness can be examined using "conditional statistical parity" [xlviii]. This approach accounts for the merit of individuals within different groups. When there are known differences in the merits, the model allows corresponding differences in the outcomes. The size of these differences is determined by statistical tests on some of key features describing the relevant merits. For example, in recruitment, statistical parity can be conditioned on features such as skills, education, and relevant experience. The test expects similar scores for individuals with a similar level of qualifications. Then the results reveal the impact of other unknown or undesired factors on the outcome, such as the tone of the language in the resumé.

Since conditional statistical parity accounts for merits, it can deal with unequal base rates among groups. Unlike simple statistical parity, it permits a perfect model as a viable option. It is also compatible with, and a complement to, the measures described in section 2 on "equal treatment".

### 3.2 Equal Accuracy

The notion of equal accuracy (also known as equalized odds) requires that AI solutions work the same way for all groups. Specifically, the difference between predicted and actual outcomes should be similar between groups. This is achieved by defining, measuring, and comparing prediction error, and ensuring that the type and the level of errors are the same (or similar) across groups.

To measure and compare error rates, organizations must select a specific definition of accuracy. Depending on the use case, accuracy can be defined in various ways. The most common definitions used for equal accuracy are rates of false positives or false negatives. It is possible to compare and monitor both errors across groups. However, AI models (or other processed) cannot be tuned to minimize both errors at the same time [xlix, l] (for more information, see section 4.1 below on the mutual incompatibility of fairness).

In addition to the definition of accuracy, the notion of equal accuracy requires reliable "actuals" data to be used as a benchmark in error calculation. This data is not always available. Many of the challenges around evaluating training data (described previously) apply to the actuals data used to measure accuracy. Although these issues may be less important when tuning the model for accuracy, they are important in evaluating fairness – particularly when the perceived actuals are influenced by the output of the AI process itself. When this occurs, there is a risk of a self-reinforcing feedback loop and bias in the data. (For more information, see the examples of feedback loops described above.)

There are also many AI use cases that do not have reliable actuals. Perhaps the most common application of AI in HR is matching candidates' resumés with job descriptions. This is a high-impact application where the evaluation of fairness is necessary. However, in such cases, there are no actuals or benchmarks available to show correct matches. This means that the equal accuracy approach has limited applicability in this case.

When applicable, equal accuracy is a robust and powerful notion of fairness. It is quantifiable and verifiable. It is compatible with the principles of data science and AI development. In fact, the models can be tuned to produce similar error rate across groups. This can be achieved by co-optimizing accuracy and bias (defined as the gap in the error rates between different groups). It ensures that the solution works the same way for all groups, and is compatible with the notion of equal treatment. In most cases, AI designers can seek both notions of equal treatment and equal accuracy at the same time. Most importantly, equal accuracy does not suffer from the flaws of statistical parity. It can deal with unequal base rates between groups and, unlike statistical parity, it permits the perfect model as a valid option.

## 4. FAIRNESS TRADE-OFFS

### 4.1 Mutual Incompatibility of Fairness

There is an inherent incompatibility among different notions of fairness. Generally, equal treatment and equal outcome cannot be satisfied at the same time. When there is a difference in group qualifications, equal treatment leads to a corresponding difference in outcomes (violating statistical parity). Similarly enforcing equal outcomes would require favoring one group over another, violating the principle of equal treatment.

Incompatibility also exists within the different notions of equal outcome (statistical parity vs. equal accuracy). Achieving equal accuracy does not lead to statistical parity and vice versa. Consider credit risk modeling, where different groups have varying historical default rates on their loans. Achieving equal accuracy on the risk of default requires different scores, and generating equal scores translates into unequal errors for the groups.

This incompatibility exists even within the same notion of fairness (e.g., equal accuracy), depending on which accuracy metrics are used. The most common metrics used for this purpose are false positive and false negative rates. There is mathematical proof that, except in highly constrained special cases, equal rates for these two metrics cannot be achieved simultaneously [li, lii].

Because of these incompatibilities, identifying the proper notion of fairness for any application requires human judgment. The pros and cons of each approach described here should be used as a guide for this purpose. Practically, many of the measures for equal treatment can be used along with equal accuracy or conditional statistical parity. The focus of equal treatment is on the evaluation and validation of the input data, mitigating the causes of bias at the source. While equal accuracy can be achieved through co-optimizing accuracy and error in the AI algorithm, for any given input data. In fact, the benchmark data used to measure accuracy should be evaluated using the same methods described in equal treatment (e.g., evaluated for risk of bias, unbalanced representation, or self-reinforcing feedback loops). These considerations make a strong case for a holistic approach to examining data used for AI modeling.

On the other hand, it is not possible to reconcile the incompatibilities of statistical parity with other notions (conditional statistical parity, equal accuracy or equal treatment). Statistical parity inherently relies on a different principle to equal treatment. The only scenario where these two approaches are compatible is where all groups have the same distribution of features or qualifications. As described earlier, this is not realistic. Statistical parity is also at odds with equal accuracy. These approaches anticipate different patterns in outcomes and seek to verify these patterns through quantitative, statistical tests. Naturally, the tests cannot be satisfied for both.

### 4.2 Fairness vs. Accuracy Trade-Offs

Fairness is not an independent aspect of AI. As outlined in section 2 above on "equal treatment", certain measures can improve both the accuracy and fairness of AI solutions. Examples include scrutinizing training data for the risk of bias, self-reinforcing feedback loops, or unbalanced representation. Other measures, such as controlling the features in the data, may compromise predictive accuracy. In this case, the impact of each feature on model accuracy can be isolated and quantified, helping AI designers understand the cost and benefits of including certain features in their models. The combined effect of the measures for equal treatment may increase or decrease overall predictive accuracy, but it would likely lead to improved accuracy for under-represented or historically disadvantaged individuals.

Demanding equal outcomes across groups, on the other hand, will always come at the cost of reduced predictive accuracy [liii, liv]. Optimizing AI models for the outcome of certain groups negatively impacts other groups. This reduction in predictive accuracy should not be taken lightly, because it results in a model rejecting qualified individuals and accepting unqualified ones. Therefore, accuracy not only impacts the solution's utility, but also leads to unfair outcomes at the individual level. Consider a randomized outcome which is unfair to qualified candidates yet delivers equal outcomes (statistical parity) at a group level.

Enforcing statistical parity in a realistic environment, with unequal base rates among groups, requires the violation of the equal treatment principle. This can be done through various methods. Since this notion does not require an AI designer to understand the causes of differences between groups, they may leave the causes unchanged and instead achieve the target patterns through arbitrary manipulation of their models (e.g., by removing any feature that correlates with the target groups). Such interventions could significantly impact the solution's utility and predictive accuracy, and lead to suboptimal trade-offs below the Pareto frontier (see below).

Equal accuracy requirements are typically achieved through more methodical means. For instance, an AI designer can adjust the model's core algorithms to co-optimize accuracy and fairness (error gaps). This means that the algorithms determine the highest level of accuracy that can be achieved for any given level of unfairness. These results can be plotted in the form of a convex curve, called the Pareto frontier [lv]. The Pareto frontier is a powerful tool for understanding the cost of fairness in terms of accuracy reduction. It helps AI designers find the right balance between fairness and accuracy for a given application.

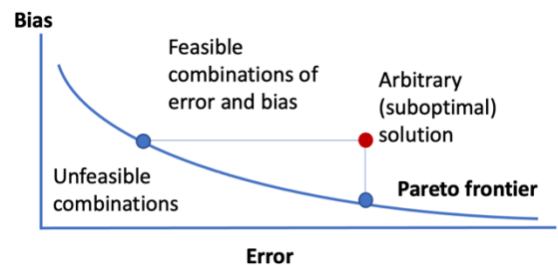

**Figure 5:** Illustration of the Pareto frontier for AI fairness.

It should be emphasized that the trade-off between accuracy and fairness always exists, even when it is not quantified and visualized in the Pareto frontier. The frontier simply reveals the reality of AI systems (or any other probabilistic decision-making system). For any given system, it is not possible to optimize the accuracy-fairness combination beyond this limit.

The Pareto frontier allows us to define the role of human judgment in setting fairness targets. While AI systems can calculate all optimum combinations of accuracy and fairness, it is the job of human decision-makers to understand the possible trade-offs, evaluate the pros and cons of each option, and make informed decisions as to which combination is a reasonable balance between these factors. In the absence of an informed decision, organizations will end up choosing an arbitrary point either on the frontier or most likely below it (i.e., a suboptimal trade-off between accuracy and fairness).

This is currently the case for the practical guidelines for adverse impact tests [lvi]. These guidelines define an arbitrary tolerance level for unfairness (typically the aforementioned four-fifths role, which permits a 20% gap in observed outcomes). In addition to other fundamental issues regarding statistical parity and the method of achieving it, arbitrary tolerances are contrary to the concept of the Pareto frontier and the realities that govern probabilistic decision-making processes.

## 5. SUMMARY AND PRACTICAL GUIDELINES

The choice of methods used to mitigate bias and evaluate fairness in AI systems depends on various factors, such as the maturity of the development team, the solution complexity, the application area, local regulations, and the stakeholders' risk tolerance. In general, it is important to focus on measures of equal treatment as the primary notion of fairness. Training data should be scrutinized to understand the presence of bias and ensure proper representation of smaller subsets within the population. These measures improve accuracy, reduce bias, address fairness at both individual and group levels, and deliver a robust solution.

When AI is developed for sensitive application areas, the solution inputs should be examined and controlled. Sensitive attributes such as demographic attributes and their proxies should be excluded from the model. Proxies should be determined using human judgment, based on their relevance to the solution's goals rather than their correlations to demographics. Models should include relevant features that describe individuals' true merit – regardless of their correlation with demographics.

Once the measures for ensuring equal treatment are in place, relevant outcome tests can be used to evaluate the gaps between the target groups. Equal accuracy is the most powerful test for outcomes. When reliable and unbiased benchmark data is available for accuracy measurement, the method has few drawbacks. The Pareto frontier concept should be adopted, along with accuracy tests to determine the acceptable level of difference between groups. When equal accuracy is applicable, no other outcomes test is necessary.

When equal accuracy is not applicable (e.g., due to a lack of reliable benchmark data), conditional statistical parity is an alternative. This method requires only the primary features used as inputs along with the final outcomes. Therefore, it can be adopted when accuracy cannot be quantified. Informed decisions should be made regarding the acceptable level of difference between groups. This is particularly important for conditional statistical parity, because it is not conditioned on all the relevant features, so some level of difference will be inevitable.

Simple statistical parity (or adverse impact tests) should only be considered as a last resort, when no other methods are applicable. This test is weak, and ineffective in correctly measuring bias in AI systems. When differences are observed using adverse impact tests, it is unclear whether they reflect different base rates among groups or indicate bias in the solution. Therefore, statistical parity can only be used as a "sufficient" and not a "necessary" condition for fairness. For all other scenarios, practitioners, organization leaders, and policymakers should transition away from adverse impact testing and adopt more sophisticated approaches.

If the AI designer's chosen outcome tests are not satisfied, they should examine the root cause of the observed difference, rather than arbitrarily tweaking their models to enforce the desired pattens. It should be reemphasized that outcome tests only provide limited visibility into the solution's fairness (e.g., the aggregated difference between certain groups included in the test). Manipulating models to achieve desired patterns for these groups will come at the expense of other groups, sub-groups, and individuals. It also leads to suboptimal solution on the Pareto frontier, and compromises the solution's accuracy and utility. In turn, lower accuracy reduces fairness at an individual level.

A methodical approach toward closing the gap between groups involves revisiting the measures of equal treatment (e.g., examination of input data, representation of groups, self-reinforcing feedback loops, and feature selection). In certain cases, and when legally permitted, the AI designer may deliberately include certain demographics in the model to close the gaps at the group level (as in the example of using commute times to predict attrition). The inclusion of demographics in the model also enables the co-optimization of fairness and accuracy. These advanced techniques should only be employed by sophisticated developers who can fully examine the impact of such attributes on accuracy and bias.

**KEYWORDS**
Artificial intelligence. AI. Responsible AI. AI ethics. Fairness.

**STATEMENTS AND DECLARATIONS**
Conflict of interest: Author works for a company that undertakes business in the deployment of AI systems as part of its commercial activities.

The opinions expressed in this document are those of the authors and do not necessarily reflect the views of Accenture.

**ACKNOWLEDGMENT**
Authoring this document would not have been possible without the support and contributions of Babak Taati (PhD), Erin Scarrow, Daniele Merico (PhD), Marisa Tricarico, Mersedeh Tariverdi (PhD), Michelle H. Keating, Poorya Ferdowsi (PhD), Pouria Lotfi, Bob Goldman, Shirleen Low, and Thomas Niven.

**FURTHER READING**
Below is a list of additional technical references (in alphabetical order) used for writing this document. These are valuable reading materials for data scientists and practitioners.

1. Cheng, P., et al.: FairFil: Contrastive Neural Debiasing Method for Pretrained Text Encoders. https://arxiv.org/abs/2103.06413 (2021)
2. Friedrich N., et al.: DebIE: A Platform for Implicit and Explicit Debiasing of Word Embedding Spaces. Cornell University. https://arxiv.org/abs/2103.06598 (2021)
3. Fukuchi, K., et al,: Prediction with Model-Based Neutrality. Springer-Verlag. https://link.springer.com/content/pdf/10.1007/978-3-642-40991-2_32.pdf (2013)
4. Grari, V., et al.: Fairness-Aware Neural Réyni Minimization for Continuous Features. Cornell University. https://arxiv.org/abs/1911.04929 (2019)
5. Gupta, P., Jaggi, M.: Obtaining Better Static Word Embeddings Using Contextual Embedding Models. Cornell University. https://arxiv.org/abs/2106.04302 (2021)
6. Kamishima, T., et al.: Fairness-aware Learning through Regularization Approach. ResearchGate. https://www.researchgate.net/publication/220766348_Fairness-aware_Learning_through_Regularization_Approach (2011)
7. Lauscher, A., et al.: A General Framework for Implicit and Explicit Debiasing of Distributional Word Vector Spaces. Cornell University. https://arxiv.org/abs/1909.06092 (2019)
8. Mary, J., et al.: Fairness-Aware Learning for Continuous Attributes and Treatments. http://proceedings.mlr.press/v97/mary19a/mary19a.pdf (2019)
9. Mehrabi, N. et al.: A Survey on Bias and Fairness in Machine Learning https://www.cs.purdue.edu/homes/bb/2020-fall-cs590bb/docs/Survery_of_Bias_and_Fairness_in_ML.pdf (2022)
10. Miroshnikov, A., et al.: Wasserstein-based fairness interpretability framework for machine learning models. Cornell University. https://arxiv.org/abs/2011.03156 (2021)
11. Rathore, A., et al.: VERB: Visualizing and Interpreting Bias Mitigation Techniques for Word Representations. Cornell University. https://arxiv.org/abs/2104.02797 (2021)
12. Rozado, D.: Wide range screening of algorithmic bias in word embedding models

using large sentiment lexicons reveals underreported bias types. Cornell University. https://arxiv.org/abs/1905.11985 (2020)
13. Tramer, F., et al.: FairTest: Discovering Unwarranted Associations in Data-Driven Applications. Cornell University. https://arxiv.org/abs/1510.02377 (2016)
14. Valentim, I., et al.: The Impact of Data Preparation on the Fairness of Software Systems. Cornell University. https://arxiv.org/abs/1910.02321 (2019)
15. Zhoa, C., Chen, F.: Rank-Based Multi-task Learning for Fair Regression. Cornell University. https://arxiv.org/abs/2009.11405 (2020)

[i] Eitel-Porter, R., Corcoran, M., Connolly P.: Responsible AI: from principles to practice. Accenture. https://www.accenture.com/us-en/insights/artificial-intelligence/responsible-ai-principles-practice (2021). Accessed 12 January 2022

[ii] Accenture: AI ethics & governance. Accenture. https://www.accenture.com/us-en/services/applied-intelligence/ai-ethics-governance. Accessed 3 June 2022.

[iii] Responsible Artificial Institute: Adopting Responsible AI in Practice. https://cdt.ca.gov/wp-content/uploads/2021/06/Adopting-Responsible-AI-in-Practice.pdf (2021). Accessed 12 January 2022

[iv] Gates, S.W., Perry, V.G., Zorn, P.M.: Automated underwriting in mortgage lending: Good news for the underserved? ResearchGate https://www.researchgate.net/publication/239749502_Automated_underwriting_in_mortgage_lending_Good_news_for_the_underserved (2002)

[v] Hall, P., Curtis, J., Pandey, P.: Machine Learning for High-Risk Applications. O'Reilly Media, Sebastopol (2022)

[vi] Yates, S. The Ethics of Affirmative Action. Foundation for Economic Education. https://fee.org/articles/the-ethics-of-affirmative-action/ (1994). Accessed 14 February 2022

[vii] Fullinwider, R. Affirmative Action. Stanford Encyclopedia of Philosophy. https://plato.stanford.edu/entries/affirmative-action/#ContEnga (2001). Accessed 14 February 2022

[viii] Mintz, S. Evaluating the Ethics of Affirmative Action Policies on University Campuses. Ethics Sage. https://www.ethicssage.com/2014/07/evaluating-the-ethics-of-affirmative-action-policies-on-university-campuses.html (2014). Accessed 14 February 2022

[ix] Executive Order 11246 – Equal Employment Opportunity. Office of Federal Contract Compliance Programs. https://www.dol.gov/agencies/ofccp/executive-order-11246/ca-11246. Accessed 14 February 2022

[x] Affirmative Action Frequently Asked Questions. Office of Federal Contract Compliance Programs. https://www.dol.gov/agencies/ofccp/faqs/AAFAQs (2021). Accessed 14 February 2022

[xi] Section 2000e-2 - Unlawful employment practices. 148 Analyses of this statute by attorneys, Casetext. https://casetext.com/statute/united-states-code/title-42-the-public-health-and-welfare/chapter-21-civil-rights/subchapter-vi-equal-employment-opportunities/section-2000e-2-unlawful-employment-practices/analysis?citingPage=1&sort=relevance&sortCiting=date-descending. Accessed 14 February 2022

[xii] Hand, J.: Positive Discrimination In Employment Under The Equality Act 2010 – The First Judicial Consideration. University of Portsmouth. https://www.port.ac.uk/news-events-and-blogs/blogs/law/positive-discrimination-in-employment-under-the-equality-act-2010#:~:text=Positive%20Discrimination%2C%20which%20entails%20treating,of%20the%20Equality%20Act%202010. (2020). Accessed 14 February 2022

[xiii] Seiner, J.: Disentangling Disparate Impact and Disparate Treatment: Adapting the Canadian Approach. 25 Yale L. & Pol'y Rev. 95 (2006-2007)

[xiv] Furnish, H. A.: A Path through the Maze: Disparate Impact and Disparate Treatment under Title VII of the Civil Rights Act of 1964 after Beazer and Burdine. 23 B.C. L. Rev. 419 (1981-1982)

[xv] Brake, D.L.: The Shifting Sands of Employment Discrimination: From Unjustified Impact to Disparate Treatment in Pregnancy and Pay. 105 Geo. L.J. 559 (2016-2017)

[xvi] Davis, J., Purves, D., Gilbert, J. et al. Five ethical challenges facing data-driven policing. AI Ethics. https://doi.org/10.1007/s43681-021-00105-9 (2022)

[xvii] Predictive Policing. Google Scholar. https://scholar.google.com/scholar?hl=en&as_sdt=0%2C9&as_vis=1&q=predictive+policing&btnG=. Accessed 22 January 2022

[xviii] Adler, P. et al.: Auditing Black-box Models for Indirect Influence. Cornell University.

[xix] Fish, B., Kun, J., Lelkes, A.D.: A Confidence-Based Approach for Balancing Fairness and Accuracy. https://arxiv.org/abs/1601.05764 (2016). Accessed 14 February 2022

[xx] Delua, J.: Supervised vs. Unsupervised Learning: What's the Difference? IBM. https://www.ibm.com/cloud/blog/supervised-vs-unsupervised-learning (2021). Accessed 14 February 2022

[xxi] Kearns, M., Roth, A.: The Ethical Algorithm, p78. Oxford University Press, Oxford (2019)

[xxii] Calders, T. et al.: Controlling Attribute Effect in Linear Regression. ResearchGate. https://www.researchgate.net/publication/261637367_Controlling_Attribute_Effect_in_Linear_Regression (2013). Accessed 14 February 2022

[xxiii] Siblini, W. et al.: Master your Metrics with Calibration. Cornell University. https://arxiv.org/abs/1909.02827 (2020). Accessed 14 February 2022

[xxiv] Zhoa, C., Chen, F.: Rank-Based Multi-task Learning for Fair Regression. Cornell University. https://arxiv.org/abs/2009.11405 (2020). Accessed 14 February 2022

[xxv] Datta, A., Swamidass, J.: Fair-Net: A Network Architecture For Reducing Performance Disparity Between Identifiable Sub-Populations. Cornell University. https://arxiv.org/abs/2106.00720 (2021). Accessed 14 February 2022

[xxvi] Roy, A. et al.: Multi-Fair Pareto Boosting. Cornell University. https://arxiv.org/abs/2104.13312 (2021). Accessed 14 February 2022

[xxvii] Faculty. Bias identification and mitigation in decision-making algorithms. Centre for Data Ethics and Innovation (2020). Accessed 12 January 2022

[xxviii] Berk, R. et al.: Fairness in Criminal Justice Risk Assessments: The State of the Art. Cornell University. https://arxiv.org/abs/1703.09207 (2017). Accessed 14 February 2022

[xxix] Zemel R., et al.: Learning Fair Representations. University of Toronto. https://www.cs.toronto.edu/~toni/Papers/icml-final.pdf (2013). Accessed 14 February 2022

[xxx] Factsheet: EU rules on gender-neutral pricing in insurance. European Commission. https://ec.europa.eu/commission/presscorner/detail/en/MEMO_12_1012 (2012). Accessed 14 February 2022

[xxxi] Bakakar, N.: Behind the Wheel, Women Are Safer Drivers Than Men. New York Times. https://www.nytimes.com/2020/04/27/well/live/car-accidents-deaths-men-women.html (2020). Accessed 14 February 2022

[xxxii] Barocas, S., Hardt, M.: NIPS 2017 Tutorial on Fairness in Machine Learning. https://fairmlbook.org/tutorial1.html (2017). Accessed 14 February 2022

[xxxiii] Racial Profiling Definition. ACLU. https://www.aclu.org/other/racial-profiling-definition. Accessed 14 February 2022

[xxxiv] Clary, J.: Racial Profiling Studies in Law Enforcement: Issues and Methodology. Minnesota House of Representatives Research Department. https://www.house.leg.state.mn.us/hrd/pubs/raceprof.pdf (2000). Accessed 14 February 2022

[xxxv] Dwork, C. et al.: Fairness through awareness. ACM Digital Library (2012). https://dl.acm.org/doi/10.1145/2090236.2090255

[xxxvi] Kearns, M., Roth, A.: Forbidden Inputs. The Ethical Algorithm, p66. Oxford University Press, Oxford (2019)

[xxxvii] Equal Employment Opportunity Commission, Office of Personnel Management, Department of Justice, Department of Labor and Department of Treasury: Adoption of Questions and Answers to Clarify and Provide a Common Interpretation of the Uniform Guidelines on Employee Selection Procedures. U.S. Government. https://www.eeoc.gov/laws/guidance/questions-and-answers-clarify-and-provide-common-interpretation-uniform-guidelines (1979). Accessed 12 January 2022

[xxxviii] Office of the Comptroller of the Currency: Fair Lending. U.S. Government. https://www.occ.treas.gov/topics/consumers-and-communities/consumer-protection/fair-lending/index-fair-lending.html. Accessed 22 January 2022

[xxxix] U.S. Equal Employment Opportunity Commission: Who is protected from employment discrimination? U.S. Government. https://www.eeoc.gov/employers/small-business/3-who-protected-employment-discrimination. Accessed 22 January 2022.

[xl] Gov.uk: Discrimination: your rights. UK Government. https://www.gov.uk/discrimination-your-rights. Accessed 22 January 2022

[xli] Justice Laws Website: Canadian Human Rights Act. Government of Canada. https://laws-lois.justice.gc.ca/eng/acts/h-6/fulltext.html (1985). Accessed 22 January 2022.

[xlii] Equal Credit Opportunity Act. Federal Trade Commission.

[xlii] https://www.ftc.gov/enforcement/statutes/equal-credit-opportunity-act. Accessed 14 February 2022

[xliii] Title IX Of The Education Amendments Of 1972. U.S. Department of Justice. https://www.justice.gov/crt/title-ix-education-amendments-1972 (1972). Accessed 14 February 2022

[xliv] The Fair Housing Ace. U.S. Department of Justice. https://www.justice.gov/crt/fair-housing-act-1#:~:text=The%20Fair%20Housing%20Act%20prohibits%20discrimination%20in%20housing%20based%20upon%20religion.&text=The%20number%20of%20cases%20filed,as%20race%20or%20national%20origin. Accessed 14 February 2022

[xlv] Equal Employment Opportunity Commission, Office of Personnel Management, Department of Justice, Department of Labor and Department of Treasury: Adoption of Questions and Answers to Clarify and Provide a Common Interpretation of the Uniform Guidelines on Employee Selection Procedures. U.S. Government. https://www.eeoc.gov/laws/guidance/questions-and-answers-clarify-and-provide-common-interpretation-uniform-guidelines (1979). Accessed 12 January 2022

[xlvi] Kearns, M., Roth, A.: Accounting for Merit. In Kearns, M., Roth, A.: The Ethical Algorithm, p72. Oxford University Press, Oxford (2019)

[xlvii] Equal Employment Opportunity Commission, Office of Personnel Management, Department of Justice, Department of Labor and Department of Treasury: Adoption of Questions and Answers to Clarify and Provide a Common Interpretation of the Uniform Guidelines on Employee Selection Procedures. U.S. Government. https://www.eeoc.gov/laws/guidance/questions-and-answers-clarify-and-provide-common-interpretation-uniform-guidelines (1979). Accessed 12 January 2022

[xlviii] Corbett-Davies, S., et al: Algorithmic Decision Making and the Cost of Fairness (2017)

[xlix] Chouldechova, A.: Fair Prediction with Disparate Impact: A Study of Bias in Recidivism Prediction Instruments. Mary Ann Liebert, Inc. πhttps://www.liebertpub.com/doi/abs/10.1089/big.2016.0047 (2017)

[l] Kleinberg, J., Mullainathan, S., Raghavan, M.: Inherent Trade-Offs in the Fair Determination of Risk Scores. Cornell University. https://arxiv.org/abs/1609.05807 (2016). Accessed 22 January 2022

[li] Chouldechova, A.: Fair Prediction with Disparate Impact: A Study of Bias in Recidivism Prediction Instruments. Mary Ann Liebert, Inc. https://www.liebertpub.com/doi/abs/10.1089/big.2016.0047 (2017)

[lii] Kleinberg, J., Mullainathan, S., Raghavan, M.: Inherent Trade-Offs in the Fair Determination of Risk Scores. Cornell University. https://arxiv.org/abs/1609.05807 (2016). Accessed 22 January 2022

[liii] Berk, R., et al: Fairness in Criminal Justice Risk Assessments: The State of the Art. Cornell University. https://arxiv.org/abs/1703.09207 (2017). Accessed 14 February 2022

[liv] Faculty. Bias identification and mitigation in decision-making algorithms. Centre for Data Ethics and Innovation (2020). Accessed 12 January 2022

[lv] Kearns, M., Roth, A.: No Such Thing as a Fair Lunch. In Kearns, M., Roth, A.: The Ethical Algorithm, p78. Oxford University Press, Oxford (2019)

[lvi] Equal Employment Opportunity Commission, Office of Personnel Management, Department of Justice, Department of Labor and Department of Treasury: Adoption of Questions and Answers to Clarify and Provide a Common Interpretation of the Uniform Guidelines on Employee Selection Procedures. US Government. https://www.eeoc.gov/laws/guidance/questions-and-answers-clarify-and-provide-common-interpretation-uniform-guidelines (1979). Accessed 12 January 2022